\documentclass[3p,times]{elsarticle}

\makeatletter
\def\ps@pprintTitle{%
 \let\@oddhead\@empty
 \let\@evenhead\@empty
 \def\@oddfoot{}%
 \let\@evenfoot\@oddfoot}
\makeatother

\usepackage{amssymb}
\usepackage{amsmath}
\usepackage{bbm}
\usepackage{booktabs}
\usepackage{multirow}
\usepackage{url}
\usepackage{listings}
\usepackage{natbib}
\biboptions{sort&compress}

\usepackage[absolute]{textpos}
\usepackage{color}
\definecolor{mygreen}{rgb}{0,0.6,0}
\definecolor{dkgreen}{rgb}{0,0.6,0}
\definecolor{gray}{rgb}{0.5,0.5,0.5}
\definecolor{mauve}{rgb}{0.58,0,0.82}

\newcommand{\idx}[1]{
    \mathrm{#1}
}

\newcommand{\hateq}{
     \hat{=} 
}

\newcommand{\diff}[1]{
     \mathrm{d} #1 
}
\newcommand{\T}{
    ^\mathrm{T}
}

\newcommand{\vect}[1]{
    \boldsymbol{#1}
}

\newcommand{\mat}[1]{
    \boldsymbol{#1}
}

\newcommand{\snr}{
    \mathrm{SNR}
}

\newcommand{\db}{
    \mathrm{dB}
}

\begin{document}

\begin{frontmatter}

\title{Nonlinear system identification employing automatic differentiation}

\author {Jan Schumann-Bischoff}
\ead{jan.schumann-bischoff@ds.mpg.de}
\author{Stefan Luther} 
\ead{stefan.luther@ds.mpg.de}
\author{Ulrich Parlitz}
\ead{ulrich.parlitz@ds.mpg.de}

\address{Max Planck Institute for Dynamics and Self-Organization \\
Am Fa\ss berg 17, 37077 G\"ottingen, Germany}
\address {Institute for Nonlinear Dynamics, Georg-August-Universit\"at
G\"ottingen \\
Am Fa\ss berg 17, 37077 G\"ottingen, Germany}
\address{DZHK (German Center for Cardiovascular Research), partner site
G\"ottingen, \\
and Heart Research Center G\"ottingen, D-37077 G\"ottingen, Germany}

\begin{abstract}
An optimization based  state and parameter estimation method is presented where
the required Jacobian matrix of the cost function is computed via automatic
differentiation. Automatic differentiation evaluates the programming code of the
cost function and provides exact values of the derivatives.
In contrast to numerical differentiation it is not suffering from approximation
errors and compared to symbolic differentiation it is more convenient to use,
because no closed analytic expressions are required. Furthermore, we demonstrate
how to generalize the parameter estimation scheme to delay differential
equations, where estimating the delay time requires attention.
\end{abstract}


\end{frontmatter}

\begin{textblock}{165}(168,262)
\footnotesize{\textit{\today}}
\end{textblock}

\section{Introduction}
\label{sec:introduction}
\begin{textblock}{165}(23,270)
\noindent \small{\copyright \text{ } 2013 Elsevier. This manuscript version is 
made available under the CC-BY-NC-ND 4.0 license 
\url{http://creativecommons.org/licenses/by-nc-nd/4.0/}. The following article 
appeared in J. Schumann-Bischoff \textit{et al.}, Commun Nonlinear 
Sci Numer Simulat \textbf{18}, 2733 (2013) and may be found at 
\url{http://dx.doi.org/10.1016/j.cnsns.2013.02.017}.}
\end{textblock}
For many processes in physics or other fields of science mathematical models
exist (in terms of differential equations, for example), but not all state
variables are easily accessible (measurable) and proper values of model
parameters may be (partly)  unknown. 
In particular, detailled biological cell models (e.g., cardiac myocytes 
\cite{CBCDFMPSSZ11})  may include many variables which 
are difficult to access experimentally and, in addition, depend on up to hundreds of physical parameters 
whose values have to be determined.
To estimate unobserved variables (as a
function of time) and model parameters different identification methods have
been devised
\cite{PJK96,VTK02,GB08,ACJ08,SO09,SRL09,YCCLP07,YP08,G08,FTNL10,CGA08,QBCKA09,
ACFK09,
QA10,AKW10,B10,BS11,SBP11} . These methods have  in common that an attempt is
made to adjust the model output (in general a function of the state variables)
to some (experimentally) observed time series. To achieve
agreement, unobserved variables and unknown model parameters are suitably
adjusted such that the model reproduces and follows  the observed time series.
In geosciences and meteorology (e.g., whether forecasting) this procedure
is often called  \textit{data assimilation} and describes the process of incorporating new 
(incoming) data into a computer model of the real system.

A general framework for state estimation provides, for example, 
the path integral formalism including a saddle point approximation \cite{A09,QA10}.
This formalism can be used to state the estimation problem as an 
optimization problem \cite{SBP11,CGA08,QBCKA09,ACFK09,B10}.
If  an optimization method is employed that is based on gradient descent
(such as the well-known Levenberg-Marquard method \cite{L44,M63}), in general
the Jacobian matrix of
the cost function has to be provided, whose derivation may be quite cumbersome
(and error-prone), depending on the structure of the cost function and the
underlying mathematical model of the dynamical system. To estimate the Jacobian
matrix one may
approximate it by numerical derivatives (often spoiled by unacceptably large
truncation errors) or use symbolic mathematics, which requires, however, that
the function to be derived has to be given in closed form. 

A convenient
alternative to both of these methods is \textit{automatic differentiation}
\cite{autodiff} where exact numerical values of the required derivatives are
computed 
by analyzing a given source code implementation of the cost function. As will be
shown here automatic differentiation leads in this context not only
to a very flexible and efficient algorithm for computing the required Jacobian 
but also provides the sparsity
pattern of the Jacobian which is exploited by suitable optimization methods. 
In Section \ref{sec:method} we will give a formal description of the
optimization problem to be solved for state and parameter estimation.
Then we briefly present in Section \ref{sec:autodiff} the concept of 
automatic differentiation in the form used here. 
As an illustrative example we show in Section \ref{sec:lorenz96} how to
estimate
the model parameters of the Lorenz-96 model. In Section \ref{sec:estimtau} we
discuss how to estimate the delay time in delay differential
equations and provide in Section \ref{sec:mackeyglass} an example (Mackey-Glass
model).

\section{State and parameter estimation method}
\label{sec:method}
The method used here to adapt a model to a time series is based on
minimizing a cost function and was introduced in Ref. \cite{SBP11}. 
For completeness we present in the following an extended version
covering also delay differential equations (DDEs).

We assume that a multivariate $R$-dimensional time series $\{\vect{\eta}(n)\}$
is given consisting of $N+1$
samples $\vect{\eta}(n) \hateq \vect{\eta}(t_n) \in \mathbb{R}^R$ measured at
times ${\cal{T}} = \{ t_n = n\cdot \Delta t \mid n = 0,1, \dots, N
\}$.
For simplicity the observation times $t_n$ are equally spaced (with a fixed time
step $\Delta t$) and start at
$t_0 = 0$. The estimation method can easily be extended to nonuniformly sampled
observations (see Ref. \cite{SBP11}). Here we consider the
general case of a model given by a set of coupled delay differential equations
(DDE)
\begin{align}
    \frac{\diff{\vect{y}(t)}}{\diff{t}} &=
\vect{F}(\vect{y}(t), \vect{y}_\tau(t), \vect{p}, t)\, ,
    \label{eq:dde_model}
\end{align}
with $\vect{y}_\tau(t) = \vect{y}(t-\tau)$.
The state vector(s) $\vect{y}(t) = (y_1(t),\dots ,
y_{D}(t))\T$, the delay parameter $\tau \in \mathbb{R}$
and the $U$ model parameters $\vect{p} = (p_1,\dots ,p_U)\T$
are unknown and have to be estimated from the time series $\{ \vect{\eta}(n)
\}$. Estimating $\tau$ can not be conducted as estimating $\vect{p}$, because
$\vect{F}(\vect{y}(t), \vect{y}_\tau(t), \vect{p}, t)$ does \textit{not}
explicitly depend on $\tau$. In fact $\vect{F}(\vect{y}(t), \vect{y}_\tau(t),
\vect{p}, t)$ depends on $\vect{y}_\tau(t)$ which is a function of $\tau$. 
We shall later come back to this topic.

Note that \eqref{eq:dde_model} also describes (as a special case) models
given by coupled
ordinary differential equations (ODEs). In this case the right-hand side of
\eqref{eq:dde_model} is independent of $\vect{y}_\tau(t)$ and thus can be
replaced by $\vect{F}(\vect{y}(t),\vect{p},t)$ (see Ref. \cite{SBP11} for
details). 

To estimate the unknown quantities a measurement function
\begin{align}
    \vect{z}(t)     &= \vect{h}(\vect{y}(t),\vect{q},t) 
    \label{eq:measurementfct}
\end{align} 
is required to represent the relation between model states $\vect{y}(t)$  and
the $\vect{z}(t)$ corresponding to the observations 
$\{\vect{\eta}(n)\}$. This measurement function may contain  $V$ additional
unknown
parameters $\vect{q}=(q_1,\dots, q_V)\T$ that also have to be estimated using
information from the given time series $\{ \vect{\eta}(n) \}$.
\subsection{Cost function}
\label{sec:cost}
The goal of the estimation process is to find a set of values for all unknown
quantities such that the model equations
provide via measurement function \eqref{eq:measurementfct} a model times series
$\{ \vect{z}(t_n) \}$ that matches the experimental time series
$\{\vect{\eta}(t_n) \}$. In other words, the average difference between
$\vect{\eta}(t_n)$ and $\vect{z}(t_n)$ should be small.
Furthermore, the model equations should be fulfilled as well as possible. This
means that modeling errors $\vect{u}(t)$ are allowed, but should be small.
Therefore, model \eqref{eq:dde_model} is extended to
\begin{align}
    \frac{\diff{\vect{y}(t)}}{\diff{t}} &=
\vect{F}(\vect{y}(t), \vect{y}_\tau(t), \vect{p}, t) + \vect{u}(t) \, .
    \label{eq:modelu}
\end{align} 

The smaller $\vect{u}(t)$ is the better the model equations \eqref{eq:dde_model}
are fulfilled. Next, for simplicity, $\vect{u}(t)$ and $\vect{y}(t)$ will be
discretized at the times in ${\cal{T}}$. This means that $\vect{y}(t)$ will be
sampled at the same time when data are observed. With 
$\vect{y}(n) \hateq \vect{y}(n\cdot \Delta t) \hateq \vect{y}(t_n)$ and 
${\cal{Y}}(a,b) = \{\vect{y}(n) \mid n =a,a+1, \dots, b\}$ the set of
values of the discretized model variables can be
written as ${\cal{Y}}(0,N)$. The quantities in 
${\cal{Y}}(0,N)$ have to be estimated in addition to $\vect{p}$ and $\vect{q}$.
With the same
discretization we have $\{\vect{u}(n)\} = \{\vect{u}(t_n)\}$. At this point we
assume a fixed (not to be estimated) delay $\tau = k \cdot \Delta t$ with $k \in
\mathbb{N}$ which is not necessarily equal to the delay parameter of the
physical process underlying the data. This simplifies the discretization
of the delayed variable to $\vect{y}_\tau(t) = \vect{y}(n\cdot\Delta t-k\cdot
\Delta t) = \vect{y}((n-k)\cdot\Delta t) = \vect{y}(n-k) = \vect{y}_k(n)$.
The set of
the discretized delayed variable is then ${\cal{Y}}_k(0,N) =
{\cal{Y}}(-k,N-k)$. 
Note that ${\cal{Y}}(-k,N-k) = {\cal{Y}}(-k,-1) \cup
{\cal{Y}}(0,N-k)$. Since ${\cal{Y}}(0,N-k) \subset {\cal{Y}}(0,N)$,
${\cal{Y}}(0,N-k)$ contains no additional quantities to be determined. Only the
variables in
${\cal{Y}}(-k,-1)$ are additional quantities which have to be estimated.
Typically the delay time is much shorter than the length of the time series
$N\cdot \Delta t$ and hence the 
number of elements in ${\cal{Y}}(-k,-1)$ is much smaller than in
${\cal{Y}}(0,N)$. Therefore the number of quantities to be estimated does not
increase much compared to a model given by ODEs (with similar $D$ and $N$)
where ${\cal{Y}}(-k,-1)$ has not to be estimated. 

The discretization of \eqref{eq:modelu} is then given by
\begin{align}
    \vect{u}(n) &\approx \left. \frac{\Delta \vect{y}}{\Delta t} \right|_{t_n}
                            - \vect{F}(\vect{y}(n), \vect{y}_k(n) ,
                            \vect{p},t_n)  \, ,
    \label{eq:findiffmodel}
\end{align}
whereas the symbol $\left. \frac{\Delta \vect{y}}{\Delta t} \right|_{t_n}$
stands for the finite difference approximation of
$\frac{\diff{\vect{y}(t)}}{\diff{t}}$ at time $t_n$. The goal of the adaption
process is to minimize (on average) the norm of $\vect{u}(n)$ \textit{and} the
norm of the difference $\vect{\eta}(t_n) - \vect{z}(t_n)$.

This leads to a cost function 
\begin{align}
    C({\cal{Y}}(-k,N),\vect{p},\vect{q}) 
    &= C_1 + C_2 + C_3 + C_4
    \label{eq:cost}
\end{align}
with

\begin{align}
    C_1 &=  \frac{\alpha}{N}\cdot\sum_{n=0}^N\left(\vect{\eta}(n)
-            \vect{z}(n)\right)\T \vect{A} \left(\vect{\eta}(n) -
            \vect{z}(n)\right) 
            \label{eq:ts_model_diff} \\
    C_2 &=  \frac{1-\alpha}{N}\cdot\sum_{n=0}^N \vect{u}(n)\T \vect{B}
            \vect{u}(n) 
            \label{eq:modellingerr} \\
    C_3 &=  \frac{1-\alpha}{N}\cdot\sum_{n = 3}^{N-2}
            \left(\vect{y}_\idx{apr}(n)-\vect{y}(n)\right)\T \vect{E}
            \left(\vect{y}_\idx{apr}(n)-\vect{y}(n)\right) 
            \label{eq:hermite4cost} \\
    C_4 &=  \frac{\beta}{L}\cdot\vect{q}
            (\vect{w},\vect{w}_\idx{l},\vect{w}_\idx{u})\T \cdot     
            \vect{q}(\vect{w},\vect{w}_\idx{l},\vect{w}_\idx{u}) \, .
            \label{eq:costbounds}
\end{align}
$C_1$ penalizes the difference between $\vect{\eta}(n)$ and
$\vect{z}(n)$ whereas $C_2$ penalizes large magnitudes of $\vect{u}(n)$. 
$A$, $B$, and $E$ are weight matrices that will be specified later.
At the minimum of \eqref{eq:cost} the solution
$(\hat{{\cal{Y}}}(-k,N),\hat{\vect{p}},\hat{\vect{q}})$ is obtained which
is considered as the solution of the estimation problem. In the term $C_3$ a
Hermite interpolation is performed to determine $\vect{y}_\mathrm{apr}(n)$
from neighboring points and the time derivatives which are,
according to \eqref{eq:modelu}, given by 
\begin{align}
    \vect{G}(\vect{y}(t), \vect{y}_\tau(t), \vect{p}, t) 
    &= \vect{F}(\vect{y}(t),\vect{y}_\tau(t), \vect{p}, t) + \vect{u}(t) \, .
    \label{eq:G}
\end{align}
With Eq. \eqref{eq:G} the Hermite interpolation reads
\begin{align}
    \vect{y}_\idx{apr}(n)  
    =&  \frac{11}{54} \left[ \vect{y}(n-2)+ \vect{y}(n+2) \right] +   
        \frac{8}{27} \left[ \vect{y}(n-1) + \vect{y}(n+1)\right]
        \notag \\
     &  + \frac{\Delta t}{18}   
        \left[
        \vect{G}(\vect{y}(n-2),\vect{y}_k(n-2),\vect{p},t_{n-2})
        -\vect{G}(\vect{y}(n+2),\vect{y}_k(n+2),\vect{p},t_{n+2})     
        \right] \notag \\
    &   + \frac{4\Delta t}{9} \left[
        \vect{G}(\vect{y}(n-1),\vect{y}_k(n-1),\vect{p},t_{n-1}) 
        - \vect{G}(\vect{y}(n+1),\vect{y}_k(n+1),\vect{p},t_{n+1})
        \right] \, .
        \label{eq:hermite4}
\end{align} 

Smoothness of ${\cal{Y}}(0,N)$  is enforced by
small differences $\vect{y}_\idx{apr}(n)-\vect{y}(n)$.
The term $C_3$ suppresses non-smooth (oscillating) solutions which
may occur without this term in the cost function. Let
\begin{align}
    \vect{w} &= ({\cal{Y}}(0,N),{\cal{Y}}(-k,-1), \vect{p},\vect{q}) \notag \\
             &= ({\cal{Y}}(-k,N), \vect{p},\vect{q}) \notag \\
             &= (w_1,\dots, w_L)
    \label{eq:w}
\end{align}
be a vector containing all quantities to be estimated \footnote{Here we assume
that a fixed but arbitrary rule is used to order the elements of the sets 
${\cal{Y}}(0,N)$ and ${\cal{Y}}(-k,-1)$ to define the elements of the vector
$\vect{w}$.}. Again, if the model is
given by ODEs, ${\cal{Y}}(-k,-1)$ does not occur in $\vect{w}$. Hence for ODEs
we
obtain
\begin{align}
    \vect{w} &= ({\cal{Y}}(0,N), \vect{p},\vect{q}) \, .
\end{align}
To force
$\vect{w}$ to stay between the lower and upper bounds $\vect{w}_\idx{l}$ and
$\vect{w}_\idx{u}$, respectively,
$\vect{q}(\vect{w},\vect{w}_\idx{l},\vect{w}_\idx{u}) = (q_1,\dots ,q_L)\T$
is defined as
\begin{align}
    q_i(w_i,w_{\idx{l},i},w_{\idx{u},i}) &= \begin{cases}
        w_{\idx{u},i}-w_\idx{i}     & \mathrm{for} \quad w_i \geq w_{\idx{u},i}
\\
    0                       & \mathrm{for} \quad w_{\idx{l},i} < w_\idx{i} <
w_{\idx{u},i} \\
      w_{\idx{l},i}-w_\idx{i}   & \mathrm{for} \quad w_i \leq w_{\idx{l},i}
\quad .\\
                                \end{cases}     
\end{align}
$q_i$ is zero if the value of $w_i$ lies within its bounds. To enforce this,
the positive parameter $\beta$
is set to a large number, e.g. $10^5$. In this paper the matrices $\mat{A}$,
$\mat{B}$ and $\mat{E}$ are diagonal matrices. The diagonal elements
can be used for an individual weighting. 

The homotopy parameter $\alpha$ can be
used to adjust whether the solution should be close to data ($\alpha \approx 1$)
or
have a smaller error in fulfilling the model equations (see Ref. \cite{B10}).
In
\cite{BS11} a possible technique is described to find an optimal $\alpha$.
Furthermore
one might use continuation (see Ref. \cite{B10}) where $\alpha$ is stepwise
decreased. Starting with $\alpha \approx 1$ results in a solution close to the
data. Then, $\alpha$ is slightly decreased and the previously obtained
solution is used as an initial guess and the cost function is be optimized
again. This procedure is repeated until the value $\alpha=0.5$ is reached.

Note that the cost function can be written in the form
\begin{align}
    C(\vect{w}) = \sum_{j=1}^J H_j(\vect{w})^2 
                = \left\Vert \vect{H}(\vect{w}) \right\Vert_2^2 
    \label{eq:costopt}
\end{align}
where $\vect{H}(\vect{w})$ is a high dimensional vector valued function of
the high dimensional vector $\vect{w}$. To optimize \eqref{eq:costopt} we use an
implementation of the Levenberg-Marquardt algorithm \cite{L44,M63}
called \texttt{sparseLM} \cite{sparseLM}. Although $C(\vect{w})$ will be
optimized, \texttt{sparseLM} requires $\vect{H}(\vect{w})$ and the sparse
Jacobian of $\vect{H}(\vect{w})$ as input. In the next section we discuss how to
derive the Jacobian and its sparsity structure.

\section{Automatic differentiation}
\label{sec:autodiff}
The technique used here to estimate the variables and the parameters of a model
from time series is based on minimizing the cost function \eqref{eq:cost} which
can be written in the form of Eq. \eqref{eq:costopt}. The
Levenberg-Marquardt
algorithm used to minimize this cost function needs the vector valued
function $\vect{H}(\vect{w})$ and its Jacobian $\partial
\vect{H} / \partial \vect{w}$ with the elements $\partial
H_j / \partial w_l$ as input. Remember that the Jacobian has a sparse
structure, i.e. it has many elements which are always zero.
To compute this sparse Jacobian there exist three different techniques:
\textit{numerical differentiation}, \textit{symbolic
differentiation} and \textit{automatic differentiation} (symbolic
differentiation includes differentiation by hand). 
These techniques have particular advantages and disadvantages.
\textit{Numerical differentiation} is easy to implement, but
numerically not exact.  Furthermore, the sparsity pattern can not be
detected reliably. 
\textit{Symbolic differentiation} is numerically exact, but the
function to be differentiated has to be available as a single expression.
Deriving the Jacobian by hand usually is very error prone. Symbolic
differentiation tools may help at this point, however, a 
change in the cost function requires deriving a new Jacobian. 
As an alternative the concept of \textit{automatic differentiation} can be used.
It is easy to implement, numerically exact and the sparsity pattern of the
Jacobian can be detected automatically. Only the source code of the cost
function is required by the automatic differentiation tool. Using automatic
differentiation requires additional computational resources.
  
After weighing up the pros and cons of the discussed methods
for computing the Jacobian we came to the conclusion that the concept of
automatic differentiation is the most suitable one. Automatic differentiation
is used here in terms of the tool (library) ADOL-C \cite{adolc,GJU96}. ADOL-C
provides functions to derive the numerical values of the Jacobian
$\partial \vect{H} / \partial \vect{w}$ of the function
$\vect{H}(\vect{w})$. Furthermore the sparsity pattern of
$\partial \vect{H} / \partial \vect{w}$ can be detected and
the numerical values of the non-vanishing elements can be derived (this functionality
requires the graph coloring package ColPack \cite{colpack}).

We used the Python interface Pyadolc \cite{pyadolc} to ADOL-C
wrapping functions of the ADOL-C library to Python. The advantage
of using the
Python interface instead of the C interface is that the cost function can
be coded directly in Python using Numpy \cite{numpy} arrays. ADOL-C
is based on operator overloading. This means, that for
computing the Jacobian the function to be differentiated
has to be available as source code, only, with an input $\vect{w}$ and return
$\vect{H}(\vect{w})$. For evaluating the cost function usually the elements of
$\vect{w}$ have a numerical data type (e.g. integer, float, ...). 

Typically deriving the sparsity pattern takes much more time than computing the
non-zero values. However, this is more or less negligible because the detection
of the sparsity pattern only has to be performed once, whereas the computation
of the values of the non-zero elements occurs several times (always when the
Jacobian has to be computed for a certain input $\vect{w}$, usually in each
iteration of a numerical minimization routine). 

For the examples in Sections \ref{sec:lorenz96} and \ref{sec:mackeyglass} the
time needed by ADOL-C
used for computing the Jacobian of Eq. \eqref{eq:costopt} was compared with
the time needed by the used minimization routine (sparseLM) and the results are
shown in Tab.
\ref{tab:adolc_times}. 
\begin{table}
    \centering
    \begin{tabular}[t]{lrr}
        \toprule 
        \textit{Example:} & Mackey-Glass model & 80-dim Lorenz96 system \\
        \textit{Section} &   \ref{sec:mackeyglass}  &   \ref{sec:lorenz96} \\
        \midrule        
        \textit{CPU time for:} & &\\
        \cmidrule{1-1} 
        cost function calls & 1.07s & 31.8s \\        
        Jacobian calls (ADOL-C)& 2.34s & 155s \\         
        optimizer (sparseLM) & 3.14s & 21845s \\         
        \bottomrule        
     \end{tabular}
     \caption{  CPU times needed for evaluation of the values of the sparse
                Jacobian compared to the CPU time needed by the optimization
                routine. For the Mackey-Glass model only the CPU time for
                $\tau \in [2.3,2.4]$ from the \textbf{second step} of estimating
                the delay time is shown. The time values were measured on a
                1.3GHz Dual core computer with 3GB memory.}
     \label{tab:adolc_times}
\end{table}     
It turned out that the time needed by ADOL-C is much shorter than the time
needed by sparseLM, i.e. the time needed to evaluate the cost function is
negligible.
Hence using ADOL-C does not lead to a significant increase of CPU time needed to
solve the estimation problem.

\section{Lorenz-96 model}
\label{sec:lorenz96}
As described in Section \ref{sec:method} the estimation method can be used
to adapt a system of ODEs to a time series (without time delay). As our first
example we use the
Lorenz-96 model that was introduced by E. Lorenz in 1996 \cite{lorenz96}. Here
the $D=80$ dimensional system is used given by the set of ODEs
\begin{align}
    \frac{\diff{y}_i}{\diff{t}} &= y_{i-1}(t)\cdot
            (y_{i+1}(t)-y_{i-2}(t))-y_i(t) + p
    \label{eq:lorenz96_model}
\end{align}
whereas $i=1, \dots, D$ is a cyclic index. This means that for $i=D$ it is
$\vect{y}_{i+1}=\vect{y}_1$. For $i=1$ it is $i-1=D$ and $i-2=D-1$. To compare
the results from the
estimation process a twin experiment is performed. The time
series $\{ \vect{\eta(t_n)} \}$ with $t_n \in \{ 0,0.01,0.02,\dots, 10 \}$ is
generated by integrating a similar model 
\begin{align}
    \frac{\diff{x}_i}{\diff{t}} &= x_{i-1}(t)\cdot
            (x_{i+1}(t)-x_{i-2}(t))-x_i(t) + 8.17
    \label{eq:lorenz96_ts}
\end{align}
with the same dimension and cyclic index and taking the solution
to build a noisy $R=D/2=40$ dimensional multivariative time series by
``observing'' every second
model variable,
\begin{align}
    \vect{\eta}_\mathrm{ts}(t_n)    &= (x_1(t_n),x_3(t_n),x_5(t_n),\dots
x_{D-1}(t_n)) \, .
\end{align}    
A common type of noise in experimentally observed time series is white noise
which is given by normally distributed random numbers with a variance $\sigma^2$
and a mean which is zero. Adding the noise to the clean time
series we obtain a noisy multivariate time series 
\begin{align}
    \vect{\eta}(t_n)    &= \vect{\eta}_\idx{ts}(t_n) + \vect{r}(t_n),
    \qquad \vect{r}(t_n) = (r_1(t_n), r_2(t_n), \dots, r_R(t_n))
    \quad \mathrm{with} \quad r_i(t_n) \sim \mathcal{N}(0,\sigma^2) 
    \label{eq:mnoise}
\end{align}
which is typical for experiments with measurement noise.
To quantify the power of the clean signal and the noise one can define the
\textbf{s}ignal-to-\textbf{n}oise-\textbf{r}atio ($\snr$ in dB) for each time
series as
\begin{align}
    \snr(\{\eta_i(t_n)\})    &= 10\cdot \log_{10}\left( \frac{\sum_{n=0}^{N+1}
                            \left(\eta_{\idx{ts},i}(t_n) -                     
                   \overline{\eta}_{\idx{ts},i}\right)^2}{\sum_{n=0}^{N+1}
                            r_i(t_n)^2} \right)                          
    \label{eq:snr}
\end{align}
(the overbar denotes the mean). The smaller the
$\snr$ the more measurement noise is present.

\begin{figure}[htp]
    \centering
    \includegraphics[width = 13.9cm] {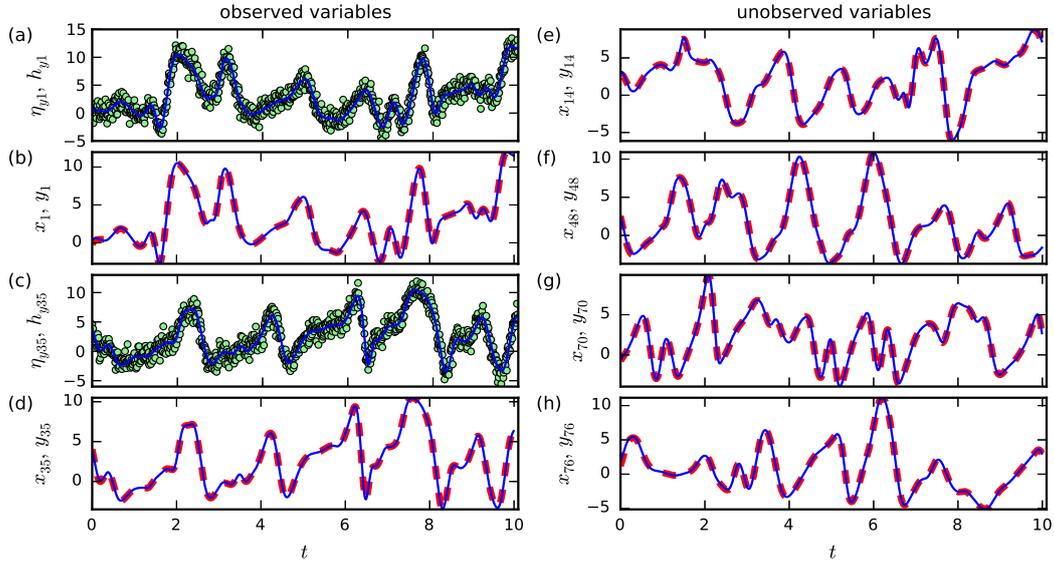}
    \caption{Adaption of the 80 dimensional Lorenz96 model (Eq.
            \eqref{eq:lorenz96_model}) to the time series $\{\vect{\eta}(t_n)\}$
            (Eq. \eqref{eq:lorenz96_ts}, green circles). 
            Every second model variable was observed.
            Left-hand side:
            The model output given by the measurement 
            function $\vect{h}(t)$ (Eq.  \eqref{eq:lorenz96_h}, blue lines) 
            was adapted to $\{\vect{\eta}(t_n)\}$ 
            ((a),(c)).
            (b) and (d) show the ``true'' solution $x_i$ (red dashed lines)
            with the corresponding estimates $y_i$ (blue lines),whereas for	    
	    these quantities there exist measurements $\eta_{yi}$. 
            Right-hand side: (e), (f), (g) and (h) show estimates for the
            model variables $y_i$ (blue lines)  and the ``true'' solutions 
            $x_i$ (red dashed lines). Note that
            for all cases shown on the right-hand side there exist no
            measurements. The model parameter is estimated to $p=8.165$,
            whereas $8.17$ is used for generating the data by
            \eqref{eq:lorenz96_ts}. Note that only a few of the 80 
            estimated model variables (with and without data) are shown.
            }
    \label{fig:lorenz96}
\end{figure}

In this example the observed time series is given by
\begin{align}
    \vect{\eta}(t_n)    &= (x_1(t_n),x_3(t_n),x_5(t_n),\dots x_{D-1}(t_n))
                            + \vect{r}(0,1.0) 
    \label{eq:lorenz96_eta}
\end{align}
with a mean of the $\snr$ of
$\overline{\snr}(\{\vect{\eta}(t_n)\})=1/R \cdot \sum_{i=1}^R
\snr(\{\eta_i(t_n)\}) \approx 10.7 \db$.
The measurement function is given by 
\begin{align}
    \vect{h}(\vect{y}(t),t)    &= (y_1(t),y_3(t),y_5(t),\dots y_{D-1}(t)) \, . 
    \label{eq:lorenz96_h}
\end{align}
Next, the model \eqref{eq:lorenz96_model} is adapted to  $\{ \vect{\eta}(t_n)
\}$ and the model variables and the parameter are estimated from the time
series. 
The weighting matrices of the cost function \eqref{eq:cost} were
fixed to: $\vect{A}=\mathbbm{1}_{D/2}$ ($D/2 \times D/2$ unity matrix),
$\vect{B}=\mathbbm{1}_D$, $\vect{E}=10^{5}\cdot \mathbbm{1}_D$
and $\beta=10^5$. As mentioned in Section \ref{sec:cost}, continuation with
$\alpha$ is used here. This means that $\alpha$ was set to the following
values: $0.9999,0.999,0.99,0.9,0.5$. First, the cost function was optimized with
$\alpha=0.9999$, then the resulting solution was used as the initial guess for
the next
optimization of the cost function with $\alpha=0.999$. This procedure was
repeated until $\alpha=0.5$ is reached. The solution of the optimization
problem with $\alpha=0.5$ was then considered as the solution of the estimation
problem.
The results of the estimated solution for the model variables are shown
in Fig. \ref{fig:lorenz96}.

One can see a good coincidence of the model variables
$\vect{y}(t)$ on the one hand and the ``true'' solution $\vect{x}(t)$
on the other hand. This is also the case for the observations
$\{\vect{\eta}(t_n) \}$ and
$\vect{h}(\vect{y}(t))$. Furthermore the modeling error $\vect{u}(t)$ is small.

\section{Delay differential equations and the estimation of delay parameters}
\label{sec:estimtau}
An application where the cost function becomes rather complex and automatic
differentiation turns out to be beneficial is the estimation of parameters and
states of DDEs.
How to derive the delayed variable $\vect{y}_k(n) = \vect{y}(n-k)$ 
from $\vect{y}(n)$ for
$\tau = k\cdot \Delta t$ with $k \in \mathbb{N}$ was already discussed in
Section
\ref{sec:cost}. In this case ${\cal{Y}}_k(0,N)$ can be computed from
${\cal{Y}}(0,N)$ without interpolation. However, if $k \in \mathbb{R}_+$ this
approach does not work anymore.
One has to interpolate ${\cal{Y}}(0,N)$ to approximate
${\cal{Y}}_k(0,N)$. Because $\Delta t$ will be small a linear
interpolation should be sufficient and hence is used here.

In the general case we have $\tau = (k'+l) \cdot \Delta t$ with $\tau \in
\mathbb{R}_+$, $k' \in \mathbb{N}$ and $l \in [0,1[$. Note that, due to these
restrictions on $k'$ and $l$, $k'$ and $l$ are uniquely defined for a given
$\tau$. Using this substitution for $\tau$ the delayed variable can be written
as
\begin{align}
    \vect{y}_k(n)   &= \vect{y}_\tau(t_n) \notag \\
                       &= \vect{y}(t_n-\tau) \notag \\
                       &= \vect{y}(t_n-(k'+l)\cdot \Delta t) \notag \\
                       &= \vect{y}(n \cdot \Delta t-(k'+l)\cdot \Delta t) \notag
\\
                       &= \vect{y}((n-k'+ l)\cdot \Delta t) \notag \\
                       &= \vect{y}(n-k'+ l) 
    \label{eq:delayy}
\end{align}
and with linear interpolation between $\vect{y}(n-k')$ and $\vect{y}(n-k'+1)$ we
obtain 
\begin{align}
    \vect{y}_k(n)  
    &= \vect{y}(n-k'+ l) \notag \\
    &= \vect{y}(n-k') + 
        \frac{\vect{y}(n-k'+ 1) - \vect{y}(n-k')}{\Delta t}
        \cdot l \cdot \Delta t \notag \\
    &= \vect{y}(n-k') + 
        \left[\vect{y}(n-k'+ 1) - \vect{y}(n-k') \right]
        \cdot l 
    \label{eq:lininterpol}
\end{align}
with $\vect{y}_k(n)  \in [\vect{y}(n-k'), \vect{y}(n-k'+1)]$ for 
$n = 0,1, \dots, N$.
Note, that for $l=0$ we have $\tau
= k' \cdot \Delta t$ and hence the same situation as in Section \ref{sec:cost},
were $\vect{y}_k(n) = \vect{y}(n-k')$ can be computed without an
interpolation of the model variables.

One possibility to estimate $\tau$ is discussed now. 
First, $\tau$ must be added in Eq. \eqref{eq:w} to the quantities to be
estimated.
Next, bounds for $\tau$ must be set in Eq.
\eqref{eq:costbounds}. This means that there are bounds to $k'$ so that (if the
smallest bound of $\tau$ is zero) we have $k' \in [1,K]$. 
If $\tau$ and therefore $k'$ and $l$ would be fixed and not estimated,
$\vect{w}$ (Eq. \eqref{eq:w}) would be
$\vect{w} = ({\cal{Y}}(-k',N),\vect{p},\vect{q})$. When $\tau$ will be
estimated, the number of elements in the history
${\cal{Y}}(-k',-1)$ can vary during the optimization
process due to variations in $\tau$ (performed by the optimizer) and in
$k'$. Because the number of elements of $\vect{w}$ has to stay constant during
one optimization process, the history with the largest possible number of
elements, given by ${\cal{Y}}(-K,-1)$, must be added to $\vect{w}$, which then
becomes
$\vect{w} = ({\cal{Y}}(-K,N),\vect{p},\vect{q}, \tau)$.
Estimating $\tau$ with this approach has one major disadvantage.
When the
optimizer evaluates the cost function with a certain $\tau$, first $k'$ and $l$
have to be derived from $\tau$. Next $k'$ is used as an index to define the
intervals $[ \vect{y}(n-k'), \vect{y}(n-k'+ 1) ]$ with $n=0,1,\dots N$ where to
interpolate the model variables for deriving the delayed variables 
$\vect{y}_k(n)$ with $n=0,1,\dots N$.
The cost function depends on the model
equations, the model equations depend on the delayed variable and the delayed
variable is, as shown in Eq. \eqref{eq:lininterpol}, a function of the model
variables. Only the latter ones are quantities to be estimated.
Hence the (sparse) Jacobian $\partial H_i(\vect{w}) / \partial w_j$ of
$\vect{H}(\vect{w})$ (see Eq. \eqref{eq:costopt}) depends on
the (sparse) Jacobian $\partial \vect{y}_k(n) / \partial
\vect{y}(n)$ of the delayed variables to the model variables. 
For any
specific $n$ there exist $K$ possible intervals $[ \vect{y}(n-\tilde{k}),
\vect{y}(n-\tilde{k}+ 1) ]$ with $\tilde{k} \in [1,K]$ where the linear
interpolation might be
performed although there is only the \textit{one} interval with $\tilde{k}=k'$
where it finally will be performed. 
%
%
The derivative of the linear interpolation to the model variable can be written
as $\partial \vect{y}_k(n) / \partial \vect{y}(n) = \partial
\vect{y}_{(\tilde{k}+l)\Delta t}(n) / \partial \vect{y}(n)$ with
$\tilde{k}=0,1,\dots, N$. Since only for $\tilde{k}=k'$ a linear interpolation
is performed, the relevant elements of the Jacobian matrix corresponding to
$\tilde{k}
\neq k$ are zero.
Nevertheless, these elements which are zero at this step are not necessarily
zero during
the entire optimization process. Therefore these elements must be included in
the sparsity pattern of nonzero elements, although, most of the time, they are
zero. The sparsity pattern must not change during an optimization
process. This fact would lead to a significant increase of (possible) nonzero
elements in the Jacobian and hence would remove the advantage of dealing with a
sparse Jacobian.

\begin{figure}[htp]
    \centering
\includegraphics[width = 7cm] {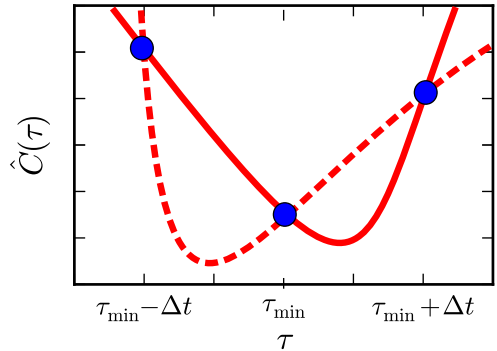}
    \caption{Estimation of the delay parameter: An illustration of the
            cost function $\hat{C}(\tau)$ around the minimum at 
            $\tau_\mathrm{min}$ is shown.
            The values of $\hat{C}(\tau)$ shown by the (blue) dots where
            obtained by the \textbf{first step} of estimating the delay
            parameter. The global minimum of $\hat{C}(\tau)$ at
            $\hat{\tau}$ is either in the interval 
            $\hat{\tau} \in [\tau_\mathrm{min} - \Delta t, \tau_\mathrm{min}]$
            or in the interval 
            $\hat{\tau} \in [\tau_\mathrm{min}, \tau_\mathrm{min} + \Delta t]$,
            as illustrated by the (red) dashed and continuous lines. Both lines
            describe possible behaviors of $\hat{C}(\tau)$ between the
            (blue) dots.
            To determine whether the global minimum is in 
            $[\tau_\mathrm{min} - \Delta t, \tau_\mathrm{min}]$ or             
            $[\tau_\mathrm{min}, \tau_\mathrm{min} + \Delta t]$ both intervals
            are investigated separately in the \textbf{second step} using
            linear interpolation of the state variable.}
    \label{fig:illustration_cost_tau_min}
\end{figure}

To avoid these problems the estimation of the delay parameter is divided into
two steps:
\begin{description}
    \item[First step:] The cost function is minimized for several fixed
        delays, $\tau = k\cdot \Delta t$ with $k \in \mathbb{N}$. In this
        case, as described in Section \ref{sec:cost}, no interpolation is
        necessary for computing the delayed variable. The solution of this
        step is $\hat{C}(\tau)$. This means that the
        assigned value is the value of the cost function in its minimum
        for a given $\tau$. $\hat{C}(\tau)$ has a global minimum 
        close to $\tau_\mathrm{min}=k_\mathrm{min} \cdot \Delta t$.
    \item[Second step:] From the first step we have values for the cost
        function \textit{only} around the minimum at $\tau_\mathrm{min}-\Delta
        t$, $\tau_\mathrm{min}$ and $\tau_\mathrm{min}+\Delta t$, as
        illustrated in Fig. \ref{fig:illustration_cost_tau_min}. The
        global minimum at $\hat{\tau}$ is either in the interval 
        $\hat{\tau} \in [\tau_\mathrm{min} - \Delta t, \tau_\mathrm{min}]$ or
        in the interval 
        $\hat{\tau} \in [\tau_\mathrm{min}, \tau_\mathrm{min} + \Delta t]$.  
        Now, $\tau$ is estimated as well. As described at the beginning of this 
        section, the delayed variable is computed by linear interpolation of
        the model variable.
        The cost function is minimized two times: once with
        $\tau \in [\tau_\mathrm{min} - \Delta t, \tau_\mathrm{min}]$   
        and then with 
        $\tau \in [\tau_\mathrm{min}, \tau_\mathrm{min} + \Delta t]$. The 
        solution (model variables, parameters, delay time)
        obtained for the interval with the smallest value of the cost 
        function after the optimization procedure is then taken as the final
        solution.
\end{description}
\section{Example: Mackey-Glass model}
\label{sec:mackeyglass}
As an example of a DDE we consider the Mackey-Glass model \cite{MG77}.
This time delay model is given by
\begin{align}
    \frac{\diff{y}(t)}{\diff{t}}    
    &= p_1 \cdot \frac{y(t-\tau)}{1+y(t-\tau)^{10}}-p_2 y(t)
    \label{eq:mg_model}
\end{align}
and has the two model parameters $p_1$, $p_2$ beside the delay parameter
$\tau$. To generate a data time series a twin experiment is performed. This
means that the model (similar to Eq. \eqref{eq:mg_model})
\begin{align}
    \frac{\diff{x}(t)}{\diff{t}}    
    &= 2 \cdot \frac{x(t-2.38)}{1+x(t-2.38)^{10}}-1\cdot x(t)
    \label{eq:mg_data}
\end{align}
was integrated first. The solution is then used to generate the (noisy) time
series
\begin{align}
    \eta(t_n) &= x(t_n) + \mathcal{N}(0,0.1)
    \label{eq:mg_eta}
\end{align}
with $t_n \in \{ 0,0.1, \dots, 60 \}$ and $N=601$ with timesteps $\Delta t =
0.1$ ($\snr = 8.0\db$ according to Eq. \eqref{eq:snr}).

\begin{figure}[htp]
    \centering
        \includegraphics{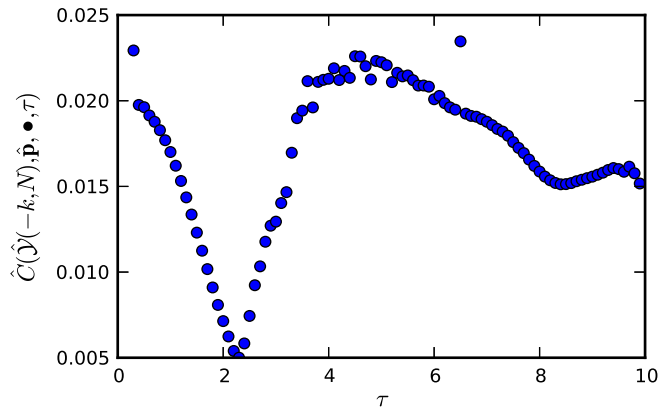}
    \caption{Cost function for adaption of the Mackey-Glass \eqref{eq:mg_model}
             model to a time
             series generated in a twin experiment with Eq. \eqref{eq:mg_eta}.}
    \label{fig:mg_cost}
\end{figure}

Next the model \eqref{eq:mg_model} was adapted to $\{ \eta(t_n) \}$ using the
measurement function
\begin{align}
    h(y(t),\bullet,\bullet) &= y(t) \, .
    \label{eq:mg_h}
\end{align}
The parameters $p_1$,
$p_2$ and $\tau$ were estimated in addition to the model variable. As described
in Section
\ref{sec:estimtau} the estimation of $\tau$ is divided into two steps:
\begin{itemize}
    \item First $\tau$ is fixed to several different values $\tau = k\cdot
          \Delta t$ with
          $k = 3, 4, \dots, 100$ and $\Delta t = 0.1$. 
          For each fixed $\tau$ the cost function is
          minimized. The value of the cost function at its minimum is
          denoted by          
          $\hat{C}(\hat{{\cal{Y}}}(-k,N),\hat{\vect{p}},\bullet,\tau)$. Its
          dependence on $\tau$ is shown in Fig. \ref{fig:mg_cost}. One can see a
          clear minimum at $\tau=2.3$. Due to the step size of $\Delta t = 0.1$
          and the smoothness around this minimum one can expect that the
          global minimum is either at $\tau \in [2.2,2.3]$ or at 
          $\tau \in [2.3,2.4]$.
    \item Second the cost function is minimized two times (first with the bounds
          $\tau \in [2.2,2.3]$ and second with $\tau \in [2.3,2.4]$).
          In each case the delayed variable is approximated by linear
          interpolation
          according to Eq. \eqref{eq:lininterpol}. Remember that due to the
          bounds the index $k'$ does \textit{not} change during each
          optimization process and hence does not have to be recomputed
          from $\tau$ in each iteration. Only $l$ changes.          
\end{itemize}
For all performed estimation processes the weighting matrices of the cost
function \eqref{eq:cost} (in this cases they are scalar) and $\alpha$ were
fixed to: $A=B=1$, $E=10^3$, $\beta=10^5$ and $\alpha=0.5$.
The results for the estimated parameters and the delay parameter are shown in 
tab. \ref{tab:estimtau}.

\begin{table}
    \centering
    \begin{tabular}[t]{@{}llll@{}}
        \toprule
        ×       & $(p_1,p_2)$ & $\tau$ & $\hat{C}(\dots, \tau)$  \\
        \midrule
        data    & $(2,1)$ & 2.38      & ×    \\
        estim. $\tau \in [2.2,2.3]$  & $(1.73,0.86)$ & 2.300 & $5.41\cdot
10^{-3}$ \\
        estim. $\tau \in [2.3,2.4]$  & $(1.98,0.99)$ & 2.387 & $4.99\cdot
10^{-3}$ \\
        \bottomrule
            \end{tabular}
 
    \caption{Model parameters $p_1$, $p_2$ and delay parameter  
             $\tau$ of the Mackey-Glass model \eqref{eq:mg_model} 
             estimated by adapting the 
             model \eqref{eq:mg_model} to
             the time series given by Eq. \eqref{eq:mg_eta}.
             According to Fig. \ref{fig:mg_cost} the cost function
             $\hat{C}(\hat{{\cal{Y}}}(-k,N),\hat{\vect{p}},\bullet,\tau)$
             has its global minimum either in $\tau \in [2.2,2.3]$ or in
             $\tau \in [2.3,2.4]$. Minimizing the cost function and estimating
             $\tau$ in addition to the model parameters and variables for both
             intervals results in a smaller
             $\hat{C}(\hat{{\cal{Y}}}(-k,N),\hat{\vect{p}},\bullet,\tau)$
             for $\tau \in [2.3,2.4]$. }  
    \label{tab:estimtau}

\end{table}

\begin{figure}[htp]
    \centering
        \includegraphics[width = 14.cm]{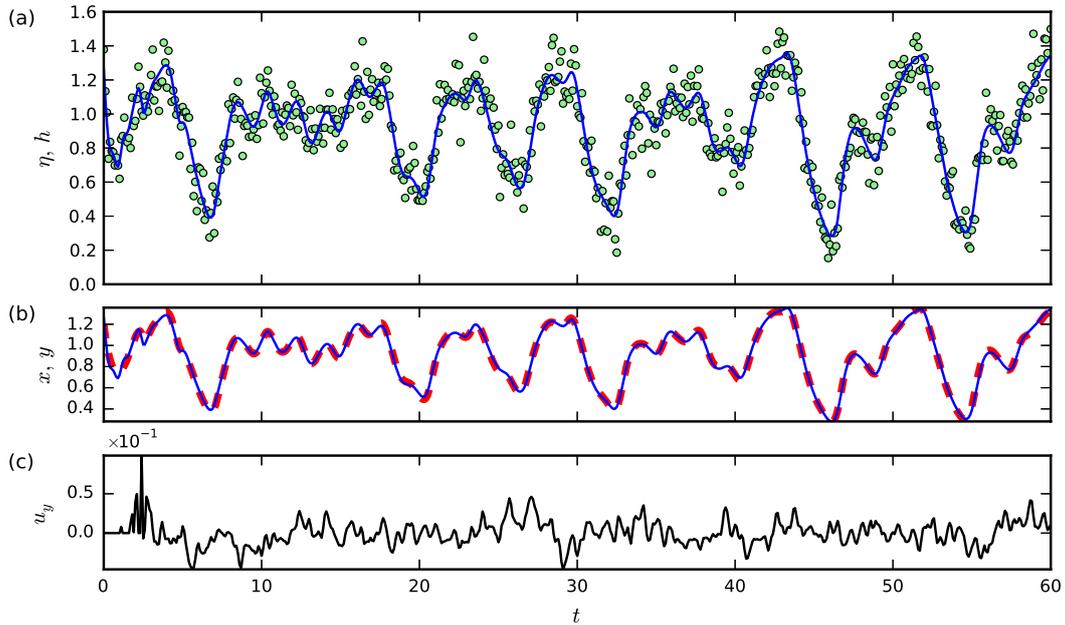}
    \caption{Adaption of the Mackey-Glass \eqref{eq:mg_model} model to the time
             series $\eta$ (see \eqref{eq:mg_eta} green circles) which was
             generated by the original ``true'' solution $x$ red dashed line,) unknown to
             the estimation algorithm) using the measurement function
             $h(y(t),\bullet,\bullet)$ (blue line, see \eqref{eq:mg_h}). 
             The estimated solution for the model variable is $y$ (blue line in (b))
             and $u$ (black line in (c)) is	 the error in the approximation
             of the model equation (see \eqref{eq:findiffmodel}). 
             The delay time
             $\tau \in [2.3,2.4]$ was also estimated beside the model
             parameters and all estimated values are shown in tab.
             \ref{tab:estimtau}.}
    \label{fig:mg_fit}
\end{figure}

For $\tau \in [2.3,2.4]$ the cost function $\hat{C}(\dots,
\tau)$ at its minimum has
a smaller value than for $\tau \in [2.2,2.3]$. Hence the solution for
$\tau \in [2.3,2.4]$ was chosen as the final result. Furthermore the estimated
values for $p_1$, $p_2$ and $\tau$ coincide much better with the ``true'' values
used to generate $\{ \eta(t_n) \}$ (compared to the estimated values with $\tau
\in [2.2,2.3]$). The estimated model variable for $\tau \in [2.3,2.4]$ is shown
in Fig. \ref{fig:mg_fit}. 
One can see a good coincidence of the model variable
$y(t)$ on the one hand and the ``true'' solution $x(t)$ (Fig.
\ref{fig:mg_fit}b) and the data
$\{ \eta(t_n) \}$ (Fig. \ref{fig:mg_fit}a) on the other hand with
a small modeling error $u(t)$ (Fig. \ref{fig:mg_fit}c). 

\section{Conclusion}
Many optimization based system identification methods require information about
 derivatives of the underlying cost function in order to converge to the desired
optimum.
In general this information has to be provided by the user (or programmer) in
terms of a Jacobian matrix, for example. Our examples show that this (typically
cumbersome) task can be conveniently handled by means of automatic
differentiation. This versatile tool from applied mathematics and computer
science not only gives exact numerical values of the required derivatives, but
also provides (and respects)
the sparsity structure of the Jacobian matrix which may be exploited by any
calling algorithm. These features have been demonstrated with a particular 
optimization algorithm (SparseLM) that enabled parameter and state estimation of
the high dimensional Lorenz-96 system and the Mackey-Glass delay differential
equation. 
A challenge for future research will be, for example, a successful application of the proposed 
parameter estimation to electrophysiological models of cardiac myocytes.

Our implementation (Python, C) of the estimation algorithm and 
a Python wrapper for sparseLM are available for download  \cite{code}.
Other possible fields of application in nonlinear dynamics are
bifurcation analysis and the computation of Lyapunov exponents and (covariant)
Lyapunov vectors \cite{KP12}.
\section*{Acknowledgements}
%
The research leading to the results has received 
funding from the European Community's Seventh Framework
Program FP7/2007-2013 under grant agreement no. HEALTH-F2-2009-241526, EUTrigTreat.
This work received support through Deutsche Forschungsgemeinschaft 
(SFB 1002 Modulatorische Einheiten bei Herzinsuffizienz).
This work was supported by the DZHK (Deutsches Zentrum f\"ur Herz-Kreislauf-Forschung - 
German Centre for Cardiovascular Research).
We thank H.D.I. Abarbanel and J. Br\"ocker for interesting and
inspiring discussions on state and parameter estimation.

\begin{appendix}
\section{Python example for automatic differentiation using ADOL-C}
    For illustration we present and discuss a Python example showing how to
    derive the sparse Jacobian of a function $\vect{H}(\vect{w})$ in Listing
    \ref{lst:pyadolcex}. 
   
   \newpage
   
    \lstset{language=Python, 
        commentstyle=\itshape\color{gray},    
        basicstyle=\footnotesize,
        numberstyle=\color{gray},
        numbersep=3pt,
        numbers=left,
        caption={Simple Python example demonstrating the derivation of the
                (sparse) Jacobian of the function $\vect{H}(\vect{w})$ usind the
                automatic differentiation tool ADOL-C.},
        label=lst:pyadolcex,
        showspaces=false,
        showtabs=false,
        showstringspaces=false,
        emphstyle={\color{self}\slshape}}
\begin{lstlisting}
import numpy as np
import adolc

def H(w):    
    h = np.zeros(3,dtype=w.dtype)
    h[0] = 3*w[0]**2 + w[1]*w[3]
    h[1] = 4*w[2]**3 
    h[2] = 5*w[0] + 2*np.exp(np.sin(w[1]*w[2]))    
    return h
    
w0 = np.array([4.,2.,3.,1.])    

# Trace H(w)
# ----------
adolc.trace_on(0)
aw = adolc.adouble(w0)
adolc.independent(aw)

Hfunc = H(aw)

adolc.dependent(Hfunc)
adolc.trace_off()

# Evaluate function jacobian at this point
w = np.array([-2.,3.,1.,-4.]) 

# Evaluate function
valH = adolc.function(0,w) 

# Evaluate dense jacobian
jacH = adolc.jacobian(0,w) 
# Out jacH: 
#   [[-12. -4.    0.     3.]
#   [  0.  0.     12.    0.]
#   [  5.  -2.280 -6.840 0.]]

# Derive sparsity pattern and evaluate values of nonzeros at w0
# -------------------------------------------------------------
options = [0, 0, 0, 0]
[nnz, rind, cind, val] = adolc.colpack.sparse_jac_no_repeat(0,w0,options)
# Out nnz (number non-zeros): 7
# Out rind (row indices): 
#   [0 0 0 1 2 2 2]
# Out cind (column indices): 
#   [0 1 3 2 0 1 2]
# Out val (values):
#   [  24. 1. 2. 108. 5. 4.356 2.904]

# Evaluate values of nonzeros at w using nnz, 
# rind, cind and valH from the  previous step
# -------------------------------------------
[nnz, rind, cind, val] = adolc.colpack.sparse_jac_repeat(0,w,nnz, rind,cind,val)
# Out nnz (number non-zeros): 7
# Out rind (row indices): 
#   [0 0 0 1 2 2 2]
# Out cind (column indices): 
#   [0 1 3 2 0 1 2]
# Out val (values):
#   [-12. -4. 3. 12. 5. -2.280 -6.840]
\end{lstlisting}

\newpage
Before ADOLC can compute the numerical values of the Jacobian data type
an internal function representation of H(w), called trace, is created in
lines 15-22. To do this the data type of the elements of the input vector is
changed to the ADOL-C data type 'adouble' (line 16). Many (mathematical)
functions are overloaded for this data type and can hence be used in the
function to be 
differentiated. FOR-loops, WHILE-loops, IF ... THEN statements, etc. are
also allowed under certain conditions.
In lines 28 and 31 the function H(w) and its dense Jacobian are computed using
the
previously created trace for a new $\vect{w}$. In line 40 the sparsity
pattern of the Jacobian is detected and used in line 52 to compute the
numerical values of its non-zero elements. The output in lines 33-35 and lines
55-59 show that the computed dense and sparse Jacobians are equal for the same
$\vect{w}$.
\end{appendix}
%





\end{document}